\documentclass{article}
\usepackage[utf8]{inputenc}
\usepackage{authblk}
\usepackage{changepage} 
\usepackage{natbib}
\usepackage{graphicx}
\usepackage[hang,flushmargin]{footmisc} 
\setcitestyle{round}

\title{Engaging Engineering Teams Through Moral Imagination: A Bottom-Up Approach for Responsible Innovation and Ethical Culture Change in Technology Companies}

\newcommand*\samethanks[1][\value{footnote}]{\footnotemark[#1]}
\author{Benjamin Lange\thanks{Joint first authors}, Geoff Keeling\samethanks, Amanda McCroskery\thanks{Creators of the program and joint second authors}, \\ Ben Zevenbergen\samethanks[2], Sandra Blascovich\thanks{Contributors to the program and contributing authors}, Kyle Pedersen\samethanks, \\ Alison Lentz\thanks{Executive supporters and contributing authors}, and Blaise Ag\"{u}era y Arcas\samethanks}

\affil{Google Research}

\providecommand{\keywords}[1]
{
  \small	
  \noindent
  \textbf{Keywords:} #1
}

\begin{document}

\maketitle



\begin{center}
    \{benlange, benzevenbergen, amccroskery\}@google.com
\end{center}

\vspace{0.2cm}

\begin{abstract}
   We propose a ‘Moral Imagination’ methodology to facilitate a culture of responsible innovation for engineering and product teams in technology companies. Our approach has been operationalized over the past two years at Google, where we have conducted over 50 workshops with teams from across the organization. We argue that our approach is a crucial complement to existing formal and informal initiatives for fostering a culture of ethical awareness, deliberation, and decision-making in technology design such as company principles, ethics and privacy review procedures, and compliance controls. We characterize some distinctive benefits of our methodology for the technology sector in particular. 

\end{abstract}

\vspace{0.5cm}


\section{Introduction}

The norms and values of technology teams shape which technologies are produced and how.\footnote{\citet{winner1980artifacts}, \citet{IdheLifeworld}, \citet{feenberg1990critical}, see also \citet{weinstein2021system}.} But these norms and values are rarely made explicit and subjected to critical appraisal, leading to limited ethical reflection and potentially reinforcing biases in technological development. In response, there are growing calls to change the culture that shapes the production of technologies. These include calls for greater governance such as government regulation and industry self-regulation by internal ethics review committees\footnote{\citet{jackman2015evolving}, \citet[p.151-57]{blackmanethicalmachines}, \citet[p.34]{shneiderman2021responsible}, \citet{prunkl2021institutionalizing}.} and governing principles.\footnote{\citet{whittlestone2019role}.} Also included are calls for engineer education in computer science curricula and industry training modules,\footnote{\citet{grosz2019embedded}, \citet{fiesler2020we}, \citet{garrett2020more}.} alongside technical best practice developments such as technical approaches to value alignment\footnote{\citet{amodei2016concrete}, \citet{kenton2021alignment}, \citet{gabriel2020artificial}.} and algorithmic auditing.\footnote{\citet{brown2021algorithm}, \citet{hasan2022algorithmic}.}

These approaches, whilst critical for promoting socially and ethically responsible technological development, have under-addressed a central cultural dynamic of technology production, namely, the group forums where critical decisions about product and research direction are made day-to-day. In the technology industry in general, these decisions are often negotiated within engineering, product, and research teams, where largely autonomous, entrepreneurially-driven groups decide which problems are addressed through technology and how, before elevating recommendations to managers and executives. To be sure, the autonomy afforded to engineering teams is a key aspect of how technology companies drive innovation. For example, in a recent study on the determinants of innovation in a Swedish software company, Jim Andersén and Torbjörn Ljungkvist found that ‘[a]lthough managers play a key role in top-down oriented innovation processes, innovation is often achieved by smaller groups at the operational level.’\footnote{\citet{andersen2021resource}.} Indeed, this observation corroborates an earlier finding by Peerasit Patanakul, Jiyao Chen, and Gary S. Lynn that ‘[r]elative to other team structures, autonomous teams are more effective in addressing projects with high technology novelty or radical innovation.’\footnote{\citet{patanakul2012autonomous}. For further empirical treatments of the role of teams and interpersonal dynamics in software engineering see \citet{scott2001strategic}, \citet{caldwell2003determinants}, \citet{karn2008ethnographic}, \citet{glynn2010fostering}, \citet{somech2013translating}, \citet{robbins2015innovating}, \citet{gerrard2018team}, \citet{hoffmann2022human}.}

In this paper, we describe a “Moral Imagination” methodology to drive a culture of responsible innovation, ethical awareness, deliberation, decision-making, and commitment in technology organizations.\footnote{Some terminological clarifications. First, we use the term “engineer” to cover a wide spectrum of roles involved in tech development, including but not limited to software engineers, data scientists, research scientists, product managers, engineering  leads, UX researchers, UX designers, or managers. We use the term ‘engineering teams’ inclusively. Second, throughout the discussion, we refer to the concept of “ethical culture” understood, roughly, as “the shared values, beliefs, norms, policies, procedures, systems, and artifacts that shape the behaviors of members of an organization and support ethical conduct.” We consider ethical culture as the encompassing construct within which initiatives such as responsible innovation or value-sensitive design that specifically aim at embedding the values or interests of broader stakeholders in technology research and development can fall.} This approach aims to prompt a \textit{role obligation shift} among teams that makes the consideration of the moral implications of their work an inherent part of their self-conception and day-to-day decision-making. As practitioners, we have developed and tested this capability building approach over the past two years at Google through over 50 workshops involving a range of research and product teams. 

Our primary aim  is to make the conceptual case for our approach, and not to present an empirical study on the efficacy of the proposed methodology, which we are pursuing in other work. Neither do we intend to suggest that our approach is superior compared to other initiatives such as traditional ethics and compliance controls, review boards, or ethics committees. Rather, we see Moral Imagination as an important complementary effort that can serve as one part of a portfolio approach to responsible innovation at technology companies.

Our discussion makes three contributions to existing scholarship on responsible innovation and technology ethics. First, it highlights a neglected gap in the current arsenal of instruments to manage responsible innovation at technology companies - the shaping of norms within engineering culture. Second, it spells out a concrete method for filling this gap, in a way that builds on existing responsible innovation frameworks, while at the same time tailoring the methodology to the distinctive cultural and organizational features of technology companies. Third, it aims to detail concrete real-world cases of application by illustrating the operationalization of this approach based on our practitioner experience.

The paper is structured as follows. Section 2 details the specific challenge that we are concerned with: equipping teams with the skills to responsibly navigate the increasing social and ethical requirements in developing their technology and products. We argue that a comprehensive culture of responsible innovation at tech companies requires interventions that work to adjust tech team norms. Against this backdrop, section 3 then develops our Moral Imagination approach. We suggest that there are three key capabilities that should be fulfilled in order to enable robust ethical culture change among teams within technology companies: Ethical Awareness, Ethical Deliberation and Decision-making, and Ethical Commitment. We then show how our framework enhances these capabilities. Section 4 concludes.

\section{The Need for Norm Shift}

\subsection{Overview}

Our argumentative strategy in this section is to show that – given a plausible assumption about the responsibility of technology companies in general – a crucial lever for enabling responsible innovation is currently not fully realized, and to then argue in the subsequent section that our approach can fill this gap. More specifically, our argument consists of the following four key claims:

\begin{enumerate}
    \item \textbf{Tech’s Practical Responsibility Requirement:} In light of the policy vacuum in which technology is developed, technology companies have a practical responsibility to consider how to produce technologies that are sensitive to the ethical and sociotechnical contexts in which those technologies will be deployed.
    \item \textbf{Importance of Shaping Norms:} Shaping engineers’ team culture and prevalent norms is a key element in responding to \textit{Tech’s Practical Responsibility Requirement}. 
    \item{\textbf{Gap in Current Measures: }Typically employed hard and soft controls are necessary but insufficient to fully shape predominant team norms in the design and development stages.} 
    \item{\textbf{New Opportunity for Intervention:} Therefore, there is an opportunity to devise new interventions that can meet the specific requirements distinctive engineering team culture poses and complement existing measures.}
\end{enumerate}

\subsection{Elaboration of Argument}

\begin{enumerate}
    \item  \textbf{Tech’s Practical Responsibility Requirement:} In light of the policy vacuum in which technology is developed, technology companies have a practical responsibility to consider how to produce technologies that are sensitive to the ethical and sociotechnical contexts in which those technologies will be deployed.
\end{enumerate}

We take this assumption as given. Our claim here draws on what has been called the “pacing problem” for technology. 

\begin{figure}[h!]
\centering
\includegraphics[scale=0.3]{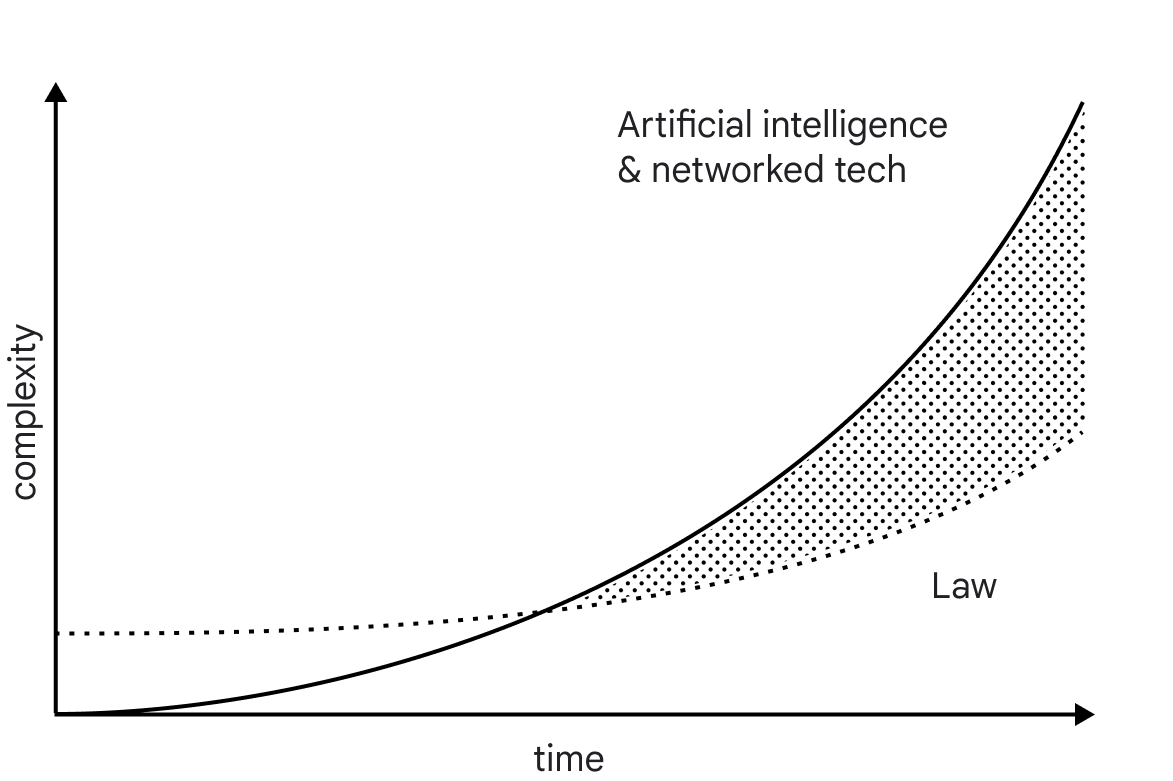}
\caption{Pacing Problem in Technology}
\label{fig:Pacing}
\end{figure}

The pace of technological innovation exceeds that of regulation.\footnote{\citet{garrett2020more}.} As a result, existing laws and policies do not necessarily provide guidance to technical companies about how to align their work with the needs and interests of society – what James Moor (1985) has also referred to as a “policy vacuum”.\footnote{\citet{moor1985computer}.} 

The pacing problem presents two \textit{ethical gray areas}. That is, ambiguities about how to proceed ethically when developing technology given that rules, laws, and policies do not straightforwardly provide guidance. The first gray area concerns the  absence of appropriate regulation for new technology, in the sense that legislation is not in place to set relevant boundaries for technological systems. The second gray area concerns the \textit{interpretation} of existing laws. This means that when regulation does exist, it has likely been designed to be appropriately broad to address foreseeable new technology. By necessity, this level of generality requires that engineers need to be able to interpret it meaningfully for the relevant context of their own work or that its interpretation may be ambiguous in the context of novel and technological developments.

Effectively navigating these ethical gray areas requires that tech companies understand and address the sociotechnical context in which their products will be deployed, in particular pre-empting various risks of harm that the technology might pose to affected stakeholders (customers, local communities, society at large, among others). 

‘Responsible Research and Innovation’ as a concept, literature, and set of processes can be seen as a response to this practical reality. Methods such as anticipatory governance,\footnote{\citet{guston2014understanding}.} technology assessment,\footnote{\citet{schot1997past}.} upstream stakeholder engagement,\footnote{\citet{wilsdon2004see}.} and value sensitive design,\footnote{\citet{friedman1996value}.} have arisen with the goal of embedding stakeholder values into the design of technology accordingly.

\begin{enumerate}
    \setcounter{enumi}{1}
    \item \textbf{Importance of Shaping Norms:} Shaping predominant engineers’ team culture norms is a key element in responding to \textit{Tech’s Practical Responsibility Requirement}. 
\end{enumerate}

We think that a key element in ensuring that engineers are adequately sensitive to the ethical and sociotechnical contexts of their deployed technology is changing the established informal rules and beliefs that govern the behavior of engineering teams. There are several reasons for this, which all concern the structural and organizational features of engineering teams, the deeply pervasive norms of engineering culture, and the agile nature of technology development.

Firstly, large technology companies like Google are, from an organizational standpoint, best understood as networks of autonomous cross-functional teams, rather than as single, aggregate entities that act in accordance with unified sets of goals.\footnote{See, for example, \citet{birhane2022values}, \citet{andersen2021resource}, and \citet{patanakul2012autonomous}.} These technology companies typically have ‘distributed’ as opposed to ‘centralized’ organizational structures that are organized around key research, products, and services. This means that teams typically possess a great deal of autonomy, and they often define, frame, and develop technology solutions before elevating recommendations to executives or senior managers. This mode of operation applies even in moments where technology companies undertake a concerted push toward a general goal.

Secondly, there are deeply ingrained norms that govern engineering culture itself. Values that relate to technical systems such as simplicity, efficiency, scalability, and elegance tend to be the focus, without explicit reference and acknowledgement of a wider set of ethical values that are also embedded in the work. Of course, there are good reasons for the cultural centrality of these values, including the ability to innovate rapidly and at scale. Engineers tend to be familiar with productive critique, optimization, and trade-offs for these types of values; however, the same know-how for negotiating these value tensions does not apply straightforwardly to social and ethical values. In practice, when individuals begin to discuss a wider set of ethically significant considerations, we find that teams are less familiar with ways to facilitate discussion about them in a productive, meaningful, and actionable way. This is also not to deny that individual engineers care a great deal about the social impact of their work; in fact we experience that engineers often consider their work as part of a larger “life project of creating good in the world.”\footnote{\citet{smith2021extracting}.} However, existing norms are often not conducive to enabling the required awareness, information-seeking, deliberation, and follow through on complex ethical issues.

Thirdly, in the early stages of technology development, there are often multiple paths that teams can pursue. Plans are actively in development, and concepts for technology research and products change rapidly, moving from a state of ambiguity into concrete formation over a period of weeks or months. In this state, the implicit beliefs, attitudes and social norms of teams exert preeminent influence over what and how technologies are built. Here social norms can determine how deliberative conversations proceed and engineers’ perceptions of which activities are necessary to shape a development process.

For  these reasons, the ethical gray areas underwrite the need for norm change in responding to tech’s practical responsibility requirement. Since engineers inevitably develop technologies in domains where the application of existing policies and laws is ambiguous or not applicable, and because they possess a great deal of autonomy in designing technology, they must – as teams be able to rely on well-developed norms of recognizing ethical issues, identifying when more information is needed, being able to reason through the different moral considerations at stake in a situation, and then practically acting on these in a way that translates them into robust commitments.

\begin{enumerate}
    \setcounter{enumi}{2}
    \item \textbf{Gap in Current Measures:} Currently available hard and soft controls are necessary but insufficient to fully shape predominant team norms in the design and development stages.   
\end{enumerate}

Technology companies typically have a large arsenal of mechanisms at its disposal to support an ethical culture and drive responsible innovation within the unique operating context of the technology sector.       

Traditionally the ethical culture of companies can be distinguished along two dimensions: formal and informal elements.\footnote{See \citet{kaptein2011understanding}.} The first, so-called “hard controls”, refer to the concrete and explicit plans, policies, and procedures within an organization. Of these formal systems, many attempt to influence company culture through intervention at an individual level. Ethics training programs are a good example of this, including the code of conduct, whistle-blowing and speak-up training, privacy and data security training, diversity, equity, and inclusion training, as well as training for responsible corporate citizenship among others.\footnote{See \citet{trevino2003managing}.} n addition to these measures, at Google there exist multiple review boards that operate at the project level, assessing research and products against Googles’ AI principles, security, and privacy standards, for example.\footnote{\citet{googleAIprinciples}.} The second element of ethical culture is informal. These so-called “soft controls” include the implicit, intangible elements, such as the values, expectations, beliefs, myths and assumptions that prevail in the organization that are not explicitly formalized through policies and processes. These informal elements also greatly matter in shaping ethical culture since the implicit norms and beliefs are key drivers of ethical conduct.

These ethics and compliance mechanisms, education and training programmes, and review boards are central to fostering a culture of responsible innovation within technology companies and our aim is not to criticize them or question their importance. However, we think that such mechanisms leave a gap with regards to the sustained promotion of a strong ethical culture in which technologies are consistently created with appropriate ethical foresight. While others have highlighted some\footnote{\citet{schiff2020principles}.} of the reasons for this apparent “principles to practices gap,”\footnote{\citet{mittelstadt2019principles}.} in our experience, a main reason for this gap is that existing hard controls do not sufficiently influence the organization at a technology team level. Engineering team norms and culture can vary somewhat widely even within a company, and  these have substantial and direct impact over ethically significant technological design decisions.

For example, while review boards, committees, and ethical commitments provide vital guardrails and checks on a  product or research against critical ethical risks of harm, they mostly come into effect at later stages of development, and thus fail to encourage explicit reflection on the ethical costs and benefits of different design strategies at earlier stages of a product development, including product ideation. Often, these stages require foundational help in developing a mindset and vocabulary for ethical analysis that enables teams to become aware of the concrete moral implications of their work, how design choices relate to trade-offs between important values and can have ethical implications for key stakeholders, how the team can then deliberate through these in an ethically sound manner, and take concrete action for the further development of their product. To do so would require alignment with a collaborative and bottom-up approach where teams build crucial ethical capabilities that are tailored to their specific issues and requirements at a fine-grained level.

Ethics trainings that are offered at an individual level encounter a different set of limitations for influencing ethically relevant design decisions amongst technology teams. While individual education can influence individuals’ beliefs, this doesn’t guarantee that the individual can successfully convince others of similar beliefs or to influence technology direction by a team. These require team solutioning and commitment. Individuals whose moral intuition has directed them to broach the topic of the ethical dimensions of their work, are confronted with fears of analysis paralysis, lack of shared understanding of vocabulary and concepts, lack of confidence in moving through difficult conversations productively, and lack of understanding about how to integrate the moral dimensions into concrete technical design decisions. Indeed, often entrenched norms “keep opinions and behaviors in place even if individuals no longer privately support them, a phenomenon known as pluralistic ignorance.”\footnote{\citet{prentice2020engineering}.}

\begin{enumerate}
    \setcounter{enumi}{3}
    \item \textbf{New Opportunity for Intervention:} Therefore, there is an opportunity to devise new interventions that complement existing measures and can meet the specific requirements that distinctive engineering team culture poses.
\end{enumerate}

What is consequently lacking is to influence which and how technologies are built, and to complement existing initiatives, are measures that directly address the culture and norms about how technology teams produce their work in the context of their work. Formats must be flexible and able to adapt to the nature of a team’s work, the various stages of their projects, and the idiosyncrasies of particular team cultures given embedded personalities and existing power dynamics. Addressing team norms directly in discussion about their work yields the opportunity to weaken existing norms and replace them with a new social contract that explicitly incorporates follow through on team responsibilities in light of agreed-upon ethical commitments.

\section{Moral Imagination}

In this section we propose a ‘Moral Imagination’ methodology that aims to promote a role-obligation shift among engineers by influencing the norms of behavior, rules, best practices, and beliefs of engineering culture at a team level. The methodology builds upon the Moral Imagination literature in business ethics, alongside ideas from the philosophy of technology and the responsible innovation literature.\footnote{For some key frameworks in the responsible innovation literature, see \citet{owen2012responsible}, \citet{stilgoe2013developing}, and \citet{van2014politics}. See \citet{werhane2008note} and \citet{werhane1999moral} on moral imagination. The \citet{fisher2006midstream} “Midstream Modulation” approach is similar in spirit to our developed method here.} We first articulate what a Moral Imagination approach amounts to and its function in the context of technology companies (3.1). We then outline a framework that specifies three key ethical capabilities around which our approach is structured (3.2). Last, we propose a method for strengthening those capabilities based on our practitioner-experience of conducting more than 40 workshops with teams at Google (3.3).

\subsection{Moral Imagination for Engineers}

We define Moral Imagination as:

\begin{adjustwidth}{0.5cm}{}
\textbf{Moral Imagination: }The ability to i) register that one’s perspective on a decision-making situation, including the available options and the normative factors relevant to adjudicating those options is limited; and to ii) creatively imagine alternative perspectives that reveal new approaches to that situation or new considerations that bear on the competing approaches.
\end{adjustwidth}

Crucial to this is “becoming aware of one’s context, understanding the conceptual scheme or “script” dominating that context, and envisioning possible moral conflicts or dilemmas that might arise in that context or as outcomes of the dominating scheme.”\footnote{\citet[p.3]{werhane2008note}} What developing Moral Imagination allows engineers to do is recognize the limitations of their pre-theoretic mental models about how their technology impacts the world, what the costs and benefits of that technology are, and what \textit{their} role is in ensuring responsible technological development.

Thus a central aim of our approach is to facilitate a \textit{role obligation shift}among engineers. It aims to shift teams’ self-conception away from a mindset where ethical considerations are removed from perceived responsibilities – something that “falls outside of the job description” – toward a mindset where the consideration of the moral implications is an inherent part of the research and development process. It aims to prompt teams to realize what they do not yet understand about how their technologies impact users, and more broadly the sociotechnical dynamics and value tensions of their technologies. It further aims to empower teams to create a map for information gathering about the issues and topics the team didn't consider before.\footnote{For a detailed exploration and description of the various elements and content of the Moral Imagination workshop see \citet{lange2023imagi}.} 

\subsection{Three Key Ethical Capabilities for Moral Imagination}

Our approach focuses on three ethical capabilities that we consider central to realizing Moral Imagination and to enhance these capabilities amongst teams to foster meaningful and productive norm change. 

What undergirds our focus on these capabilities is a conception of teams as moral group agents who have the ability to reach informed moral judgements through awareness and reasoning, act with intent, and to be held accountable for their own actions.\footnote{See \citet{rest1986moral}.} These focal points also relate to our prior discussion insofar as the reality of autonomous technology teams operating in ethical gray areas requires an enablement approach that builds teams’ ability to navigate complex ethical challenges and translate this into concrete actions and change along their product or research lifecycle.

\begin{adjustwidth}{0.5cm}{}
\textbf{Ethical Awareness:} Ability to recognize normatively significant factors and implications (e.g. moral values, ethical risks of harms, constraints and rights violations) in situations, decisions, and other relevant choice scenarios.
\end{adjustwidth}

A precondition for robust ethical deliberation, decision-making, and commitment is to expand the team’s perceptual paradigm beyond that of established engineering norms, while also sensitizing the team to moral discourse.\footnote{\citet{clarkeburn2002test}.} Developing an understanding of moral values, their normative force, action-guidingness, appropriate definitions of ethical terms for work-contexts, and how these relate to the technology and products that a team is developing are all crucial elements of this capability. In addition to shaping participants’ understanding of moral values, ethical awareness also pertains to risks of harm to various stakeholder groups, especially in a sociotechnical and not just technical context.

\begin{adjustwidth}{0.5cm}{}
\textbf{Ethical Deliberation and Decision-Making:} Ability to engage in reasoning and deliberation in relevant choice scenarios, including tensions between value and other moral commitments, conflicts, moral dilemmas, and trade-offs.
\end{adjustwidth}

Once the team has a better understanding of the ethical dimensions of their work, alongside a grasp of key ethical vocabulary, teams can be introduced to conceptual tools which allow them to understand and negotiate situations in which the competing normative considerations come into conflict. This may encompass covering conceptual distinctions concerning ‘pro-tanto’ and ‘all-things-considered’, the gradeability of normative concepts and values, including different degrees to which conflicts can occur, the notion of weighing different moral factors that may relate to a choice situation for a team in a way that is ethically rigorous and robust, and the idea that which moral factors are apparent or significant may vary based on perspective. This point is relevant because the status quo of engineering culture is often primarily consequentialist and can accordingly be broadened by being introduced to different moral considerations besides outcome-oriented utility calculations. 

\begin{adjustwidth}{0.5cm}{}
\textbf{Ethical Commitment:} Ability to derive and set concrete plans to guide further product development and/or research.
\end{adjustwidth}

Increased ethical awareness and decision-making capacities enables teams to navigate complex ethical challenges as part of their work. But building these capabilities will miss their mark if there is no commitment and accountability with respect to translating these insights into practical change. To that end, teams need to address what it means to act ethically and with integrity in their product or research context. This may mean deviating from widely accepted norms about the content, sequence, and pace of design, development, and release activities. What it means to operationalize ethical commitment varies depending on organizational structures. At Google, we co-develop with teams a set of actionable responsibility objectives that can inform Product Requirement Documents (PRD) and individual or team Objectives and Key Results (OKRs).

\subsection{Methodology}

In the previous subsection we discussed the key ethical capabilities that the Moral Imagination approach intends to influence to facilitate an ethical role obligation shift among engineering teams. In this subsection, we detail a practical four-step workshop method to strengthen these capabilities based on our experience of conducting Moral Imagination workshops with teams at Google.  

Our method expands upon existing responsible innovation frameworks such as those developed by \citet{owen2012responsible} and \citet{stilgoe2013developing}, as well as 
\citet{fisher2006midstream}, by providing a tailored methodology for facilitating responsible innovation for product and research teams that engage in software development for at-scale technologies including artificial intelligence.\footnote{Here our aim is to sketch conceptually how the workshop format fosters the relevant capabilities. We elaborate on the practical details of the workshop in a separate practice-based piece of work.}

For instance, while these approaches provide frameworks for facilitating responsible innovation in broad strokes, our approach is specifically tailored to the day-to-day realities of engineering teams in technology companies on the ground. Similarly, we adapt Werhane's (\citeyear{werhane1999moral, werhane2008note}) notion of moral imagination within this overall approach in a way that is tailored to the specific needs of engineering teams.

Our approach is operationalized through a series of workshops that are facilitated by a multidisciplinary team with academic backgrounds in ethics and/or practical experience with ethics in the technology industry. The workshop provides a structured engagement forum to assist teams typically at early stages of their work, for example, during the ideation, experimentation, prototyping, piloting, or re-imagining phases.\footnote{The benefits of engagement with innovation teams at an early to mid stage of their work has been outlined in \citet{fisher2006midstream}.} Workshops are designed to draw attention toward the salient dimensions of engineers’ work, and model and support how they can work through them together while building a shared capacity for ethical awareness, productive debate, solution finding, and planning. The workshops are specifically adapted and tailored to a team’s progress and work: they are modular and involve content that is customized for relevance to the dilemmas teams face – though it is always centrally focused on the key ethical capabilities of Awareness, Deliberation and Decision-Making, and Commitment through Moral Imagination.

The workshops employ a non-didactic approach to ethics, in the sense that the aim is not to  lecture participants about key moral principles and considerations. Nor is it to impose a particular ethical framework. Rather, our approach is to construct exercises that enable engineers to reframe their work through an ethical lens, and then re-envision their work and its corresponding responsible development process. In doing so, our goal is to align with the technology industry’s culture of autonomy and entrepreneurship, while building momentum from many engineers’ expressed desire to drive their innovations toward socially beneficial ends.

At a high level, the Moral Imagination workshops involve the following four-step structure. 

\begin{enumerate}
    \item \textbf{(Reflection)} Externalization of a team’s current moral intuitions, beliefs, and convictions about their work.  
\end{enumerate}

Norms and beliefs about a team’s work have to be made explicit to be challenged and altered. This first step therefore aims to surface and understand the particular ethical paradigm with which a team is operating in their day-to-day work by enabling teams to reflect on, articulate and clarify the values they feel are currently motivating or inherent in their work. Semi-structured discussions are used to surface the values that the team brings to bear in their work, including personal motivations, beliefs about the technological benefits, and envisioned characteristics of a world where the technology has been successfully deployed and is ubiquitous, aiming to formulate a positive vision for their technology. Building upon these discussions, facilitators introduce the concept of values in ethics, groups negotiate the most important values, and work to clarify and interpret them in the particular context of their technology. Teams also reflect on whether and in what respects current plans instantiate or fail to instantiate the stated values, and surface tensions between values that require tradeoffs.

\begin{enumerate}
    \setcounter{enumi}{1}
    \item \textbf{(Expansion)} Challenging a team’s perspective for the purpose of reflecting on their moral intuitions. Envisioning possibilities for the acquisition of ethically relevant information to inform a team’s approach, and for the work itself.
\end{enumerate}

Once teams’ moral intuitions and beliefs have been made more explicit among the group, the next step is to \textit{challenge} those intuitions and facilitate the internalization of ethical considerations beyond those that were initially surfaced. As part of building ethical awareness, the focus at this stage is to challenge the teams’ paradigm from an ethical point of view to help the group consider the key ethical implications that their work contains and also surface relevant knowledge gaps.

The centerpiece of this section is a bespoke technomoral scenario that extends the underlying logic of  each  team’s technology 5 or 10 years into the future. The scenario complicates the interplay of technology and society, ends on a cliffhanger, and emphasizes the importance of gaining different points of view as a means to anticipate ethical considerations. Participants role-play in small groups and argue the case against each other, putting to practice their ability to interpret values, argue for or against them in technological design, and build comfort with critical evaluation of their work. Throughout this section, further value tensions are solicited, documented, and described. An inclusion-focused exercise then aids participants in understanding the needs and interests of multiple stakeholder groups and how to include their voices to improve decision making. Participants are invited to a ‘veil of ignorance’ scenario where they are encouraged to envision, articulate and elaborate on the issues that might arise for the stakeholder ecosystem.\footnote{See \citet{weidinger2023using} for a recent discussion of using Rawls' veil of ignorance to align AI systems with principles of justice.} This exercise alerts participants to the possibility their team’s perspective is limited, enumerates an initial set of perspectives from which the work would benefit, and emphasizes diverse perspectives collected equitably must be a high priority. Other exercises include anticipation of sociotechnical harm, in which teams are exposed to a taxonomy of harm and brainstorm a number of concrete adverse impacts their work could potentially have, alongside alternative paths for the work in light of those possible impacts.

\begin{enumerate}
    \setcounter{enumi}{2}
    \item \textbf{(Evaluation)} Reasoning through a number of ethical perspectives about the team’s work. 
\end{enumerate}

\textit{Reflection} and \textit{Expansion} aim to build the ethical awareness of teams, specifically with an eye towards enabling a better grasp of ethically relevant factors including risks of harm. Once these have been surfaced and internalized in the context of the teams’ own technologies, the next step focuses on helping the team learn to deliberate and reason through concrete ethical choice scenarios that are relevant to them. So, after a team’s ethical paradigm has been made explicit and challenged by the team itself, ethical reasoning tools are successively introduced to enable the team to learn to reason through  trade-offs and choice scenarios that they have identified as arising in the context of their work. 

During Moral Imagination workshops at Google, moral theories are introduced schematically, and presented as a set of reasoning tools that enable participants to approach a problem from different angles. Key notions such as the “weighing” of competing moral values, trade-offs, and gradability of moral commitments are introduced to teams. Exercises then involve participants responding to arguments, formulating responses from multiple perspectives, and discussing amongst each other to reach a consensus on how best to resolve a particular value tension. The elements described aim to provide teams with a shared foundation to have ethical conversations in a pluralistic manner that goes beyond entirely deontological or consequentialist paradigms. 

\begin{enumerate}
    \setcounter{enumi}{3}
    \item \textbf{(Action)} Translation of insights and learnings into concrete team practices.
\end{enumerate}

This last step focuses on supporting teams in taking actions based on their learnings. Participants reflect on prior discussions and articulate ethical focus areas that can inform the technology concept and design in future work. Moderators work with participants during and after the workshop to shape these focus areas into responsibility objectives, which serve as actionable statements that can shape OKRs or be included in PRDs. The workshop’s focus on discussion, clarification, and negotiation ensures that the responsibility objectives are broadly supported and considered as legitimate North Stars for the team.

Moral Imagination workshops enable participants to challenge beliefs and begin to reshape the norms that guide decision-making and planning in their team. Importantly, the norm change at issue here is shared and co-constructed. Furthermore, in our experience, workshops render participants more aware of the value of seeking accurate information about how their work will function as a sociotechnical artifact, and also empower participants to interpret this information in the context of research and development processes. To that end, participants are able to proactively identify and mitigate ethical risks, which supports them in making better use of other ethics controls such as review boards, and in particular empowers teams with a degree of moral autonomy when engaging with these other ethical controls. This holistic approach, on which the Moral Imagination methodology complements more traditional hard controls, ultimately enables engineering teams to develop technologies in a way that is morally informed and which better meets the challenges for technology companies that we articulated in Section 2.

\section{Conclusion}

In this paper, we introduced the Moral Imagination approach as a method for driving ethical culture change within technology companies. The approach is a “soft control” method that emphasizes externalization and multiperspectival evaluation of the norms and values that precipitate innovation within teams through semi-structured deliberation and negotiation, alongside co-development of action-oriented ethical commitments (for example, through OKRs and PRDs). The Moral Imagination approach has been executed over 40 times at Google, and is positioned alongside “hard controls” such as ethics and privacy reviews that together make up Google’s portfolio approach to fostering a culture of responsible innovation in line with Google’s AI Principles.\footnote{\citet{pichai2018ai}.} We have argued that Moral Imagination is uniquely well-positioned to complement and address the limitations of more traditional “hard controls” in the context of technology companies, where team norms and values exhibit substantial influence over research and product decisions given the bottom-up and highly autonomous engineering culture that drives innovation within these companies. Our hope is that the Moral Imagination approach can serve as a template to foster a culture of responsible innovation across the industry.

While we are encouraged by the early results of the Moral Imagination approach, we continue to refine the approach and develop new tools and resources to scale the program within Google. This includes a dedicated empirical research track focused on measuring the efficacy of the approach as a method for ethical culture change, alongside the development of new workshop modules that aim to further upskill teams on topics such as ethical reasoning, critical reflection on metrics, and many other topics. Furthermore, we aim to contribute to and enrich the social conversation around ethical culture change within technology companies by publishing case studies alongside the findings of our empirical research, and also by externalizing the methodology to solicit participation and critical input from a broad range of stakeholders.

\bibliographystyle{plain}
\bibliography{references}

\end{document}